\documentclass{article}
\usepackage{spconf,amsmath,graphicx,hyperref}
\usepackage{multirow}
\usepackage{booktabs}
\usepackage{arydshln}

\newcommand{\myheading}[1]{%
  \addvspace{\baselineskip}
  \noindent\textbf{#1. }%
}

\title{SyncSpeech: Efficient and Low-Latency Text-to-Speech based on \\ Temporal Masked Transformer}
\name{Zhengyan Sheng$^1$, Zhihao Du$^2$, Shiliang Zhang$^2$, Zhijie Yan$^2$, Liping Chen$^{1*}$\thanks{*Corresponding author. This work was supported in part by the National Key Research and Development Program Project under Grant 2024YFE0217200.}}
\address{$^1$University of Science and Technology of China\\
   $^2$Independent Researcher\\}
\begin{document}
\ninept
\maketitle
\begin{abstract}
Current text-to-speech (TTS) models face a persistent limitation: autoregressive (AR) models suffer from low generation efficiency, while modern non-autoregressive (NAR) models experience high latency due to their unordered temporal nature. To bridge this divide, we introduce SyncSpeech, an efficient and low-latency TTS model based on the proposed Temporal Mask Transformer (TMT) paradigm. TMT synergistically unifies the temporally ordered generation of AR models with the parallel decoding efficiency of NAR models. 
TMT is realized through a meticulously designed sequence construction rule, a corresponding training objective, and a specialized hybrid attention mask. Furthermore, with the primary aim of enhancing training efficiency, a high-probability masking strategy is introduced,  which also leads to a significant improvement in overall model performance.
During inference, SyncSpeech achieves high efficiency by decoding all speech tokens corresponding to each newly arrived text token in a single step, and low latency by beginning to generate speech immediately upon receiving the second text token from the streaming input. Evaluations show that SyncSpeech maintains speech quality comparable to the modern AR TTS model, while achieving a 5.8-fold reduction in first-packet latency and an 8.8-fold improvement in real-time factor. Speech samples are available at \href{https://SyncSpeech.github.io/}{https://SyncSpeech.github.io/}.
\end{abstract}
\section{Introduction}
\label{sec:intro}

In recent years, groundbreaking innovations in generative models, along with the scaling of training data, have driven significant breakthroughs in text-to-speech (TTS) systems \cite{valle, voicebox, ns3}, enabling synthesized speech to increasingly approach the naturalness and quality of real recordings. These advancements have greatly facilitated the practical deployment of TTS technology, enhancing its impact in areas such as assistive communication \cite{fastspeech2}, virtual assistants \cite{gpt4}, and automatic video dubbing \cite{dubbing}.

Mainstream TTS models are primarily categorized into two paradigms: autoregressive (AR) and non-autoregressive (NAR).
Modern AR TTS \cite{valle}  can be formulated as a conditional language modeling task, where speech tokens are generated sequentially in a left-to-right, temporally ordered fashion.
The inherent temporal ordering  of AR TTS models lends itself naturally to a streaming generation paradigm. 
However, generation efficiency is fundamentally limited by the high frame rate of speech tokens required in this step-by-step process.  While several methods have been proposed to mitigate this issue, including group modeling in VALL-E 2 \cite{valle2}, speculative decoding in MEDUSA \cite{MEDUSA}, and codec merging in VALL-E R \cite{Valler}, these methods have yielded only modest efficiency gains. Meanwhile, advanced techniques such as denoising diffusion \cite{diffusion} and masked generative models \cite{maskgit} have been applied into the NAR TTS \cite{F5tts, maskgct}, enabling higher generation efficiency through parallel prediction.
 These NAR TTS models operate by first adding noise or applying masks based on a predicted duration.  
Subsequently, context from the entire sentence is leveraged to perform denoising or mask prediction in a temporally unordered fashion\footnote{
Here, “temporally” refers to the physical time of speech samples, not the iteration step $t \in [0, 1]$ of the above NAR models.}.
However, this non-sequential approach prevents NAR models from generating speech incrementally, resulting in high first-packet latency \cite{livespeech2}.

In addition, driven by the need to interface with upstream large language models, the focus of some AR TTS research has shifted to generating speech from streaming text. For instance, CosyVoice2 \cite{cosyvoice2.0} and IST-LM \cite{yang2024interleaved} utilize interleaved text-speech modeling to handle streaming text input. However, these models still suffer from the inherent inefficiency of AR paradigms, as they produce only a single speech token at each generation step.


To address above limitations, we propose \textbf{SyncSpeech}, 
 built upon the proposed \textbf{T}emporal \textbf{M}asked generative \textbf{T}ransformer (TMT) paradigm, which integrates the AR paradigm’s sequential modeling strengths with the parallel prediction capability of the NAR paradigm. 
 Specifically, to process streaming text input, SyncSpeech first predicts BPE-level durations via top-k sampling, conditioned on all preceding speech tokens. 
This duration prediction, along with the parallel generation of the speech tokens, is seamlessly integrated into a single decoding step. 
This unified process culminates in \textbf{text-synchronous} streaming speech generation, a paradigm whose time complexity is decoupled from the speech tokens, scaling instead linearly with the input text sequence.

Experimental results show that, while maintaining comparable speech quality, SyncSpeech significantly outperforms existing mainstream TTS models in real-time factor (RTF), achieving a 6.4 fold speedup for English and an 8.8 fold speedup for Mandarin. When integrated with LLMs, SyncSpeech further reduces first-packet latency (FPL) by factors of 3.7 and 5.8 for English and Mandarin, respectively. These findings highlight SyncSpeech’s potential as a foundational TTS model for seamless integration with upstream LLMs and deployment in latency-critical scenarios.

\setlength{\textfloatsep}{6pt}
\begin{figure*}
    \centering
    \includegraphics[width=0.8\linewidth]{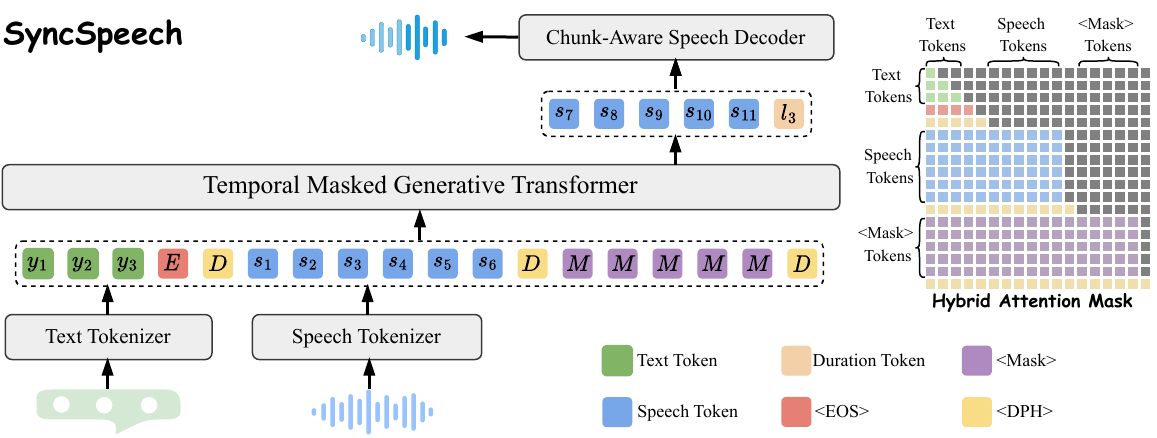}
    \caption{An overview of the proposed SyncSpeech,  comprising a text tokenizer, 
 a speech tokenizer, a temporal masked generative transformer and a chunk-aware speech decoder.
 The figure shows that, with the random number $n=2$ and text look-ahead value  $q=1$, it estimates all speech tokens (from $s_7$ to $s_{11}$) corresponding to the text token $y_2$ and the duration ($l_3$) of the next text token $y_3$ in one decoding step. The right panel illustrates the hybrid attention mask of TMT, which integrates both causal and bidirectional patterns. The special tokens $<$EOS$>$ and $<$DPH$>$ represent the end-of-text and a placeholder for duration prediction, respectively.
 }
\label{fig1}
\end{figure*}
 
\section{Method}

The overall architecture of SyncSpeech comprises two components: a text-to-token model and a token-to-speech model. As illustrated in Figure~\ref{fig1}, the proposed TMT serves as the backbone of the text-to-token model, while an off-the-shelf chunk-aware speech decoder \cite{cosyvoice2.0} from CosyVoice 2 is adopted as the token-to-speech module.

\subsection{Temporal Masked Generative Transformer}

Given a dataset of transcribed speech ($\boldsymbol{\Tilde{x}}$, $\boldsymbol{\Tilde{y}}$), where $\boldsymbol{\Tilde{x}}$ and $\boldsymbol{\Tilde{y}}$ denote an audio sample and its transcript, respectively, the transcript $\boldsymbol{\Tilde{y}}$ is tokenized into a BPE token sequence $\boldsymbol{y} = [y_1, y_2, y_3, ..., y_L]$ with a text tokenizer, where $L$ is the number of BPE tokens.  An off-the-shelf semantic speech tokenizer \cite{cosyvoice2.0} is used to encode the speech sample $\boldsymbol{\Tilde{x}}$ into $T$ frame discrete speech tokens $\boldsymbol{s} = [s_1, s_2, s_3, ..., s_T]$. We further define duration tokens $\boldsymbol{a} = [a_1, a_2, a_3, ..., a_L]$ as the positions indicating the end time of each corresponding BPE token within the speech token sequence, with $a_L = T.$  For a pair of ($\boldsymbol{\Tilde{x}}$, $\boldsymbol{\Tilde{y}}$), $\boldsymbol{a}$ can be obtained through an open-source alignment tool\footnote{https://github.com/MontrealCorpusTools/Montreal-Forced-Aligner}.

\myheading{Sequence Design} 
The sequence construction rule plays a crucial role in overall performance. To maintain consistency with the inference process (see Section Inference), a random truncation strategy is adopted during the training phase. As shown in Figure \ref{fig1}, we select a random number $n \in [1, L]$, which indicates that when receiving streaming text input, TMT needs to  generate the speech tokens corresponding to the $n$-th BPE token  at this moment. To avoid unnatural pauses, TMT allows look ahead $q$ text tokens, obtaining a truncated text token sequence $\boldsymbol{y}' = [y_1, y_2, y_3, ..., y_{L'}]$, where $L'=min(L, n+q)$. Based on the duration tokens $\boldsymbol{a}$, the truncated speech token sequence $\boldsymbol{s}_{1:a_{n}}=[s_1, s_2, ...., s_{a_{n}}]$ is obtained. Then, we define the masked speech token sequence $\boldsymbol{s}'$ and corresponding binary mask 
$\boldsymbol{m}$ as follows, 
\begin{equation}
\boldsymbol{s}' = \boldsymbol{s}_{1:a_{n}} \odot \boldsymbol{m},
\end{equation}
\begin{equation}
    \boldsymbol{m}=[m_{i}]_{i=1}^{a_{n}}, \boldsymbol{m}_{1:a_{n-1}}=0, \boldsymbol{m}_{a_{n-1}:a_{n}}=1.
\end{equation}
where $s_i$ is replaced by a special $<$MASK$>$ token if ${m}_{i} = 1$,  and otherwise leaving $s_i$ if  ${m}_{i} = 0$.
That is all speech tokens corresponding to $\boldsymbol{x}_{n}$ are replaced with the specific mask token, while the rest remain unchanged.  Then, the truncated text token sequence $\boldsymbol{y}'$, along with the masked speech token sequence $\boldsymbol{s}'$ and duration tokens $\boldsymbol{a}$, are used to construct the input sequence as follows,
\begin{equation}
\boldsymbol{f} = [\boldsymbol{y}',E, D, \boldsymbol{s}'_{1:a_1},..., D, \boldsymbol{s}'_{a_{n-1}:a_n}, D],
\label{eq1}
\end{equation}
where $E$ is end-of-text token, $D$ is a placeholder for duration prediction. 
Specifically, $D$ is used to separate the masked speech token sequence $\boldsymbol{\hat{s}}$ corresponding to different BPE tokens based on the duration tokens $\boldsymbol{a}$. 

\myheading{Loss Function} The sequence $\boldsymbol{f}$ is used as input for the TMT with the mask prediction and duration prediction as training objectives. Specifically, the sequence $\boldsymbol{f}$ is fed into the TMT to obtain the hidden states, which then pass through two different linear layers to predict the speech tokens corresponding to text token $y_{n}$ and the duration of the \textbf{next text token} $y_{n+1}$. 
 This integrates duration prediction and mask prediction into a single decoding step during inference, except for  duration prediction of the first text token.
We minimize the following negative log-likelihood function for masked generative training and duration training,
\begin{equation}
\mathcal{L}_{\text {mask}}=-\log p \left( \boldsymbol{s}_{a_{n-1}:{a_n}} \mid \boldsymbol{f}; \theta \right), 
\end{equation}
\begin{equation}
\mathcal{L}_{\text {duration}}=-\log p \left(l_{n+1}\mid \boldsymbol{f}; \theta \right),
\end{equation}
where $\theta$ represents the neural network parameters of TMT, $l_{n+1}=a_{n+1} - a_{n}$ and $a_0=0$. In this way, we simulate the scenario of receiving streaming text input, thereby enabling SyncSpeech to generate speech in sync.


\myheading{Hybrid Attention Mask} 
TMT employs a hybrid attention mask that combines both causal and bidirectional patterns, as shown in Figure \ref{fig1}. Causal attention is applied to input text tokens and special tokens. Bidirectional attention is applied to the masked and speech tokens, enabling them to attend not only to all preceding tokens but also to all masked and speech tokens corresponding to the same text token. In this way, speech tokens can perceive the total duration of their corresponding text tokens, which improve speech robustness and naturalness.
 
\myheading{Inference}
SyncSpeech can process text in a streaming fashion and generate speech  during inference.
Specifically, once the number of input text BPE tokens $\boldsymbol{y}$ exceeds the look-ahead threshold $q$, an input sequence $\boldsymbol{f} = [\boldsymbol{y}, D]$ is constructed and fed into the TMT to predict the duration of $y_1$. Based on the predicted duration, sequence padding is performed by inserting mask tokens and a duration prediction placeholder. The updated sequence is then fed back into the TMT to predict the speech tokens corresponding to $y_1$ and the duration of $y_2$, followed by an update to the input sequence $\boldsymbol{s}$ and additional padding. For each subsequent text token received, the aforementioned prediction, update, and padding steps are repeated, thereby enabling synchronous generation of each text token and all its corresponding speech tokens. During this process, once the number of generated speech tokens exceeds the chunk size of the speech decoder, these tokens and the speaker prompt can be used to generate speech waveform. It is worth noting that separate positional embeddings are employed for text and speech tokens. This design choice enables the use of a KV-cache to speed up the generation process, even when streaming text tokens are inserted.

\myheading{High-Probability Masked Pre-training}
During our exploration, we initially trained TMT from scratch, which was verified to handle streaming text input while generating natural speech with low latency and high efficiency. However, this training strategy was inefficient because at each step, the gradient was backpropagated for \textbf{only one text token}. To address this issue, we introduced high-probability masked pre-training, which not only accelerated convergence as anticipated but also led to notable gains in overall performance, an outcome not initially expected.
Here, masked speech tokens are defined as  $\boldsymbol{\hat{s}} = \boldsymbol{s} \odot \boldsymbol{\hat{m}} $, where $\boldsymbol{\hat{m}}=[\hat{m}_{i}]_{i=1}^{a_{L}} $ is a binary mask of speech tokens. 
The masking rule is designed primarily from two perspectives, high masking probability and consistency with the inference process as much as possible.
Specifically, a binary mask of text tokens $\boldsymbol{\hat{m}}_\text{bpe}$ is constructed first, where the first value is sampled according to a Bernoulli distribution ($p=0.5$) and the subsequent adjacent values cannot be the same. Based on the duration tokens $\boldsymbol{a}$, the text token mask $\boldsymbol{\hat{m}}_\text{bpe}$ is converted into the corresponding speech token mask $\boldsymbol{\hat{m}}$. Then, the input sequence is constructed in the following way, 
\begin{equation}
\boldsymbol{\hat{f}} = [\boldsymbol{y},E, D, \boldsymbol{\hat{s}}_{1:a_1},..., D, \boldsymbol{\hat{s}}_{a_{L-1}:a_L}],
\end{equation}
and the TMT is optimized to minimize the negative log-likelihood for masked generative training and duration training.
The designed hybrid attention mask mentioned above
is also utilized for pre-training. In summary, a high-probability masked pre-training is initially performed to facilitate the alignment between text and speech tokens. Subsequently, we fine-tune the pretrained model using a training strategy consistent with the prediction process. This approach enhances the efficiency of the training process, and the high-probability masked pre-training also contributes to the robustness of the generated speech.

\subsection{Other Modules}
\label{sec:other modules}
The other modules of SyncSpeech are built upon the open-source CosyVoice2\footnote{\url{https://github.com/FunAudioLLM/CosyVoice}}. Specifically, the Supervised Speech Semantic (S3) tokenizer from CosyVoice2 serves as the speech tokenizer for SyncSpeech, while its conditional flow matching decoder and HiFi-GAN vocoder \cite{hifigan} collectively form the speech decoder, which synthesizes speech waveforms from chunk-sized semantic tokens.

\section{Experiment}

\subsection{Experimental Settings}
We trained SyncSpeech on datasets of varying scales and languages, including the 585-hour English LibriTTS dataset \cite{libritts} and a 100,000-hour of internal Mandarin dataset. The open-source Montreal Forced Aligner (MFA) \cite{mfa}  aligned transcripts according to its phone set, after which the phone-level alignment was transformed into the BPE-level alignment. We evaluated SyncSpeech using three benchmarks: LibriSpeech \textit{test-clean} \cite{librispeech}, SeedTTS \textit{test-zh} \cite{seedtts}. We configured the text token look-ahead number to $q=1$ and set the speech decoder's chunk size to 15.  We use a Llama-style \cite{llama} transformer as the backbone of TMT, but replacing causal attention mask with proposed hybrid attention mask. 
The models were trained on  NVIDIA A100 80G GPUs. All training stages employed the AdamW optimizer, implementing a learning rate schedule that linearly warmed up over 32K training steps to a peak rate of $1 \times 10^{-5}$ before undergoing linear decay. To ensure a fair comparison, we re-implemented the CosyVoice AR TTS series, training it on the same dataset and using a model size, speech tokenizer, and decoder identical to those of our proposed SyncSpeech. 

\begin{table*}[t]
\centering
\setlength{\tabcolsep}{2mm}
\begin{tabular}{lcccccc}
\toprule
\textbf{Model}   & \textbf{WER(\%)} $\downarrow$   & \textbf{SS(\%)} $\uparrow$ & \textbf{MOS-N} $\uparrow$ & \textbf{FPL-A(s)}$\downarrow$ & \textbf{FPL-L(s)} $\downarrow$ & \textbf{RTF(\%)} \\ \hline
\multicolumn{7}{c}{\textbf{LibriSpeech \textit{test-clean}}}   \\ \hline
\textbf{Ground Truth} &2.12 &69.67  &$\text{4.62}_{\pm 0.12}$   \\ \hdashline
\textbf{CosyVoice}  & 3.47  & 63.52 & $\text{4.39}_{\pm 0.12}$   &0.22 &0.94 &0.45  \\ 
\textbf{CosyVoice2} & 3.00      & 63.48  &$\text{4.48}_{\pm 0.13}$ &0.22 &0.35 &0.45   \\
\textbf{SyncSpeech}  & 3.07  & 63.47 &$\text{4.48}_{\pm 0.14}$  & \textbf{0.06} &\textbf{0.11} &\textbf{0.07}  \\  \hline
\multicolumn{7}{c}{\textbf{Seed \textit{test-zh}}}   \\ \hline
\textbf{Ground Truth}   & 1.26  & 75.15 & $\text{4.68}_{\pm 0.10}$   \\ \hdashline
\textbf{CosyVoice}  & 3.63    & 72.34   & $\text{4.51}_{\pm 0.14}$   & 0.23 &0.63 &0.44          \\ 
\textbf{CosyVoice2} & 1.45      & 74.81 & $\text{4.59}_{\pm 0.13}$   &0.23 &0.36 &0.44        \\
\textbf{SyncSpeech}  & 1.43    & 74.45  &$\text{4.57}_{\pm 0.11}$    & \textbf{0.04} & \textbf{0.10} &\textbf{0.05}         \\

\bottomrule
\end{tabular}
\caption{Speech quality evaluation results of SyncSpeech and baseline models across the two benchmarks. Within the same benchmark setting, SyncSpeech and the baseline models are trained using identical model sizes and the same data scale for a fair comparison.}
\label{table1}
\end{table*}

We evaluated three key aspects across the benchmarks: speech quality, latency, and efficiency. The evaluation of speech quality was twofold, involving both objective and subjective tests. The objective metrics, aligned with CosyVoice2, included Word Error Rate (WER) and speaker similarity (SS). For the subjective evaluation, we conducted a Mean Opinion Score (MOS) test for naturalness (MOS-N), where 30 native listeners rated 100 randomly sampled sentences per system on a 1-to-5 scale.
In terms of latency and efficiency, the performance of various models was evaluated on a single NVIDIA A800 GPU.
Due to the off-the-shelf speech decoder, we evaluate the latency and efficiency of the text-to-token stage across all models.
We calculated the time required for the number of speech tokens to reach the chunk size of the speech decoder as First-packet latency (FPL). There are two scenarios, one assumes the text is already available (FPL-A), while the other involves receiving output from the upstream LLM model (FPL-L). The FPL-L of SyncSpeech, comprising the time waiting for upstream text and the time required to generate the initial speech packet, is calculated as follows:
\begin{equation}
L_{\text{FPL-L}}^{\text{SyncSpeech}} =(q+1) \cdot d_{\text{LLM}} + c \cdot d_{\text{TTS}},
\end{equation} 
where $q$ is is the number of look-ahead tokens and $c$ denotes  number of iterations needed for the generated speech tokens to exceed the decoder's chunk size. 
Here, we assume the upstream LLM model is Qwen-7B. 
When benchmarked on a single NVIDIA A800 GPU, it achieves an average token generation time, denoted as $d_{\text{LLM}} = 25 ms$. The FPL-L for other baseline models is calculated in a similar manner. When the first term in FPL-L is omitted, it becomes FPL-A. For the real-time factor (RTF), ratio of the total duration of generated entire speech to the total time taken by the model was calculated. 

\subsection{Main Results}
The evaluation results for the speech quality of SyncSpeech and the baseline models are presented in Table \ref{table1}.
SyncSpeech demonstrates speech quality comparable to the CosyVoice series, as evidenced by our evaluation across WER, SS, and MOS-N metrics. Specifically, the WER is on par with that of CosyVoice2, and the SS score is nearly identical to the CosyVoice models which is expected given the shared speech decoder.
MOS evaluations for naturalness also reveal no significant difference between the models. 

SyncSpeech demonstrates a significant breakthrough in latency. In terms of FPL-A, it is 3.7$\times$ faster than autoregressive (AR) models on LibriSpeech \textit{test-clean} and 5.8$\times$ faster on the Seed \textit{test-zh} benchmark. The greater acceleration on Mandarin is attributed to the higher compression rate, which often represents entire phrases with a single token. Furthermore, SyncSpeech excels in streaming scenarios, initiating synthesis after only two text tokens, whereas CosyVoice2 requires five and other baselines need the full sequence. This low-latency advantage is confirmed by our FPL-L evaluation. In terms of RTF, SyncSpeech is 6.4$\times$ and 8.8$\times$ faster than AR models on the LibriSpeech and Seed benchmarks, respectively. This efficiency gain stems from a fundamental shift in time complexity. While AR models operate at $O(T)$ complexity, dependent on the speech sequence length ($T$), SyncSpeech's text-synchronous paradigm achieves $O(L)$ complexity, dependent only on the text length ($L$). Since $L \ll T$ in all practical applications, this results in a decisive efficiency improvement.

{
\setlength{\textfloatsep}{0pt}
\begin{table}[]
\centering
\setlength{\tabcolsep}{1mm}
\begin{tabular}{lcc}
\toprule
      & \textbf{WER(\%)}$\downarrow$  & \textbf{UTMOSv2}$\uparrow$  \\ \hline
\textbf{SyncSpeech}   & \textbf{2.44} & \textbf{3.46}   \\ \hdashline
w/o pre-training       &  3.61 & 3.31  \\
w/o hybrid attention mask  & 8.19 & 2.98 \\ 
\bottomrule 
\end{tabular}
\caption{Evaluation results of the ablation study.}
\label{table3}
\end{table}
}

\subsection{Analysis}
We conducted ablation studies on two key components, as detailed in Table~\ref{table3}. First, removing the high-probability masked pre-training stage significantly degraded performance, increasing WER by 1.17 and decreasing the UTMOSv2 \cite{baba2024utmosv2} score by 0.15. This result underscores the critical role of pre-training in both speech robustness and naturalness. 
Second, replacing our hybrid attention mask with a conventional causal mask led to a marked decline in robustness and naturalness. This degradation occurs because causal structure prevents bidirectional attention among speech segments corresponding to a single text token. This demonstrates the the proposed hybrid attention mask, allowing the model to dynamically model the duration of each text token, was a critical factor for robust synthesis.

{
\setlength{\textfloatsep}{0pt}
\begin{table}[]
\centering
\setlength{\tabcolsep}{2mm}
\begin{tabular}{lcc}
\toprule
\textbf{Sampling Strategy}       & \textbf{WER(\%)}$\downarrow$  & \textbf{UTMOSv2}$\uparrow$  \\ \hline
\multicolumn{3}{c}{Duration Prediction} \\
\hline
Ground Truth            & 2.59 & 3.45   \\ \hdashline 
Greedy Search   & 2.50 & 3.44   \\
Top-k 3         & \textbf{2.44} & \textbf{3.46}   \\
Top-k 5         & 2.93 & 3.44   \\
Top-k 10        & 2.76 & 3.41  \\ 
\hline
\multicolumn{3}{c}{Speech Token Prediction} \\
\hline
Greedy Search & \textbf{2.44} & \textbf{3.46}   \\
Top-k 3          & 3.82 & 3.43  \\
Top-k 5          & 4.23 & 3.43   \\ 
\bottomrule
\end{tabular}
\caption{Performance across various Top-k thresholds for duration prediction and speech token prediction.}
\label{table4}
\end{table}
}

{
\setlength{\textfloatsep}{0pt}
\begin{table}[t]
\centering
\setlength{\tabcolsep}{1mm}
\begin{tabular}{cccc}
\toprule
                 \textbf{LH Num.}           & \textbf{WER(\%)}$\downarrow$  & \textbf{FPL-L(s)}$\downarrow$  & \textbf{UTMOS-v2}$\uparrow$  \\ \hline
                     q=1    & \textbf{2.44}    & \textbf{0.11}    & 3.46     \\
                     q=2    & 2.87    & 0.13   & 3.41       \\
                     q=3    & 2.52   & 0.16    & \textbf{3.48}         \\ 
                     q=4    & 2.52    & 0.19    &  \textbf{3.48}     \\ 
\bottomrule
                    
\end{tabular}
\caption{Performance with different numbers of look-ahead text tokens on LibriTTS validation set.}
\label{table5}
\end{table}
}

We analyzed the decoding strategies for duration and speech token prediction, as detailed in Table~\ref{table4}. For duration prediction, we found that Top-k=3 sampling yields superior performance in both speech robustness and naturalness. For speech token prediction, SyncSpeech exhibited a characteristic distinct from AR models: it achieves optimal performance with greedy search. We hypothesized that strict text-synchronous alignment in SyncSpeech introduces subtle temporal dependencies into the speech token sequence, rendering greedy search highly effective.

We investigated the impact of the number of text tokens to look ahead on LibriTTS validation set.
The analysis, detailed in Table~\ref{table5}, indicates that setting $q=1$ yields the best performance in terms of WER. This is primarily attributed to the effective avoidance of ambiguity and the simplification of text-to-speech alignment.
Regarding speech naturalness, a look-ahead window of $q > 2$ yielded improvements in prosody, such as more natural pacing and pauses, but this resulted in increased latency.




\section{Conclusion}
This paper presents SyncSpeech, a TTS model based on the novel TMT paradigm, which fundamentally unifies temporally-ordered generation and parallel inference. 
Our evaluations demonstrate that SyncSpeech achieves significantly improved generation efficiency and reduced latency while maintaining comparable speech quality to current AR TTS systems. 
Future work explore a unified multilingual MFA tool and integrate the proposed TMT paradigm into LLMs, ultimately developing end-to-end large speech language models.

\vfill\pagebreak

\label{sec:refs}
\bibliographystyle{IEEEbib}
\bibliography{strings,refs}

\end{document}